\documentclass[10pt,journal,compsoc]{IEEEtran}
\usepackage[T1]{fontenc}
\usepackage{amsmath}
\usepackage{amsfonts}
\usepackage{amssymb}
\usepackage{geometry}
\usepackage{booktabs}
\usepackage{enumitem}
\usepackage{hyperref}
\usepackage{parskip}
\usepackage{xcolor}
\usepackage{graphicx}  
\usepackage{float}     
\usepackage{fancyhdr}  
\usepackage{titlesec}  
\usepackage{microtype} 
\usepackage{caption}   
\usepackage{subcaption} 
\usepackage{listings}  
\usepackage{tikz}      
\usepackage{array}     
\usepackage{colortbl}  
\usepackage{longtable} 
\usepackage{adjustbox} 
\usepackage{tabularx}  
\usepackage{textcomp}  
\usepackage{multirow}  
\usepackage{natbib}
\usepackage{listings}
\usepackage{xcolor}
\usepackage{orcidlink}
\usepackage{url}
\usepackage{breakurl}

\title{Fortifying the Agentic Web: A Unified Zero-Trust Architecture Against Logic-layer Threats}

\author{Ken~Huang\,{\hypersetup{urlbordercolor=white, linkbordercolor=white}\orcidlink{0009-0004-6502-3673}},
        Yasir~Mehmood,
        Hammad~Atta\,{\hypersetup{urlbordercolor=white, linkbordercolor=white}\orcidlink{0009-0000-7801-3096}},
        Jerry~Huang, 
        Muhammad~Zeeshan Baig\,{\hypersetup{urlbordercolor=white, linkbordercolor=white}\orcidlink{0000-0002-0902-9497}},
        Sree~Bhargavi~Balija
        
\IEEEcompsocitemizethanks{
\IEEEcompsocthanksitem K. Huang is an AI Security Researcher at DistributedApps.AI, Co-Author of OWASP Top 10 for LLMs, and Contributor to NIST GenAI. (E-mail: kenhuang@gmail.com)
\IEEEcompsocthanksitem Y. Mehmood is an Independent Researcher, Germany. (E-mail: yasir.mehmood@qorvexconsulting.com)
\IEEEcompsocthanksitem H. Atta is with Qorvex Consulting \& Roshan Consulting. (E-mail: hatta@qorvexconsulting.com; hammad@roshanconsulting.ca)
\IEEEcompsocthanksitem J. Huang is with Kleiner Perkins.
\IEEEcompsocthanksitem M. Z. Baig is Course Director at Wentworth Institute of Higher Education \& Machine Learning Professional. (E-mail: muhammad.baig@win.edu.au)
\IEEEcompsocthanksitem Corresponding author is K. Huang.
\protect
 }
 {
 }}

\begin{document}
\maketitle
\begin{abstract}
This paper presents a Unified Security Architecture that fortifies the Agentic Web through a Zero-Trust IAM framework. This architecture is built on a foundation of rich, verifiable agent identities using Decentralized Identifiers (DIDs) and Verifiable Credentials (VCs), with discovery managed by a protocol-agnostic Agent Name Service (ANS). Security is operationalized through a multi-layered Trust Fabric which introduces significant innovations, including Trust-Adaptive Runtime Environments (TARE), Causal Chain Auditing, and Dynamic Identity with Behavioral Attestation. By explicitly linking the LPCI threat to these enhanced architectural countermeasures within a formal security model, we propose a comprehensive and forward-looking blueprint for a secure, resilient, and trustworthy agentic ecosystem. Our formal analysis demonstrates that the proposed architecture provides provable security guarantees against LPCI attacks with bounded probability of success.
\end{abstract}

\section{Introduction}

Autonomous AI agents are becoming the fundamental units of digital interaction, promising a new era of decentralized applications and automated workflows known as the Agentic Web. However, the underlying infrastructure for this emerging ecosystem is being built on dangerously inadequate security foundations. Traditional Identity and Access Management (IAM) protocols like OAuth, OIDC, and SAML, designed for human users and monolithic applications, fail in this new paradigm \cite{huang2025zerotrust}. Their coarse-grained, static permissions and single-entity focus cannot manage the complex delegations, ephemeral nature, and autonomous decision-making of AI agents.

This architectural gap creates profound vulnerabilities that extend far beyond conventional cybersecurity threats. Recent research has identified that AI agents introduce unique attack surfaces spanning cognitive, temporal, and operational dimensions \cite{narajala2025securing}, including risks to digital identity governance as defined in the Digital Identity Rights Framework (DIRF) \cite{atta2025dirf} and reasoning stability challenges as modeled in the Qorvex Security AI Framework (QSAF) \cite{atta2025qsaf}. The autonomous nature of these systems, combined with their ability to maintain persistent memory and execute complex reasoning chains, creates opportunities for sophisticated attacks that can remain dormant for extended periods before activation.

This paper exposes a critical flaw by defining and analyzing Logic-layer Prompt Control Injection (LPCI), a new vulnerability class where dormant, conditionally-activated malicious commands are embedded within an agent's persistent memory \cite{atta2025lpci}. These payloads can be triggered across sessions by specific events, effectively bypassing conventional defenses and turning the agent into an unwitting accomplice. Unlike traditional prompt injection attacks that target immediate response manipulation, LPCI represents a paradigm shift toward long-term, persistent system compromise that exploits the implicit trust that agentic systems place in their own memory and retrieved data sources.

Our research demonstrates that LPCI is not merely a theoretical concern but a practical and immediate threat that fundamentally challenges existing security paradigms. The existence of such attacks makes it clear that a new, purpose-built security paradigm is required that moves beyond static defenses to embrace a Zero-Trust model specifically designed for autonomous agents operating in distributed environments.

The emergence of platforms such as LangChain, LangFlow, AutoGen, and CrewAI has democratized the creation of autonomous agents, but this simplicity of integration comes with novel security threats in the form of supply chain vulnerabilities from the use of third-party modules and lightly screened components [8]. As these systems become more prevalent in enterprise environments, the need for comprehensive security frameworks becomes increasingly urgent. 

The remainder of this paper is organized as follows. Section \ref{sec:lit_review} provides a comprehensive review of related work in AI agent security, prompt injection attacks, and zero-trust architectures, and \textbf{positions the contributions of this paper in relation to prior studies}. Section \ref{sec:formal_threat_model} presents our formal mathematical model of LPCI attacks and the broader threat landscape. Section \ref{sec:archi_foundation} describes the architectural foundations of our Zero-Trust IAM paradigm. Section \ref{sec:trust_fabric} details the unified security architecture and its multi-layered Trust Fabric. Section \ref{sec:security_innovation} presents our security innovations for proactive defense against LPCI attacks. Section \ref{sec:formal_security_analysis} provides formal security analysis and evaluation of the proposed approach. Section \ref{sec:architectural_resilience} discusses architectural resilience against adaptive adversaries.Section \ref{sec:imp_consideration} discusses implementation considerations and future work. Section \ref{sec:conclusion} concludes the paper.

\section{Literature Review}
\label{sec:lit_review}

The security of AI agents represents a rapidly evolving field that intersects multiple research domains, including AI security, multi-agent systems, cybersecurity, distributed systems, and formal verification. This section provides a comprehensive review of the current state of research and positions our work within the broader landscape of AI agent security.

\subsection{Prompt Injection Attacks and Defenses}

Prompt injection attacks have emerged as one of the most significant security threats to large language models and AI systems. The fundamental vulnerability arises from the difficulty of distinguishing between legitimate instructions and malicious inputs in natural language processing systems \cite{google2025promptinjection}. Traditional prompt injection attacks focus on immediate response manipulation, where attackers craft inputs designed to override system instructions and elicit unintended behaviors.

Recent research has identified several categories of prompt injection attacks. Direct prompt injection involves crafting malicious prompts that are directly input to the system, while indirect prompt injection involves embedding malicious instructions in data sources that the AI system accesses \cite{google2025promptinjection}. The latter category is particularly concerning for agentic systems that interact with external data sources and maintain persistent memory across sessions.

Google's recent work on mitigating prompt injection attacks proposes a layered defense strategy that includes input validation, output filtering, and behavioral monitoring \cite{google2025promptinjection}. Their approach emphasizes the importance of multiple defensive layers, as no single technique can provide complete protection against the evolving landscape of prompt injection attacks. However, these defenses are primarily designed for stateless interactions and do not adequately address the persistent memory and complex reasoning capabilities of autonomous agents.

Microsoft's approach to AI security emphasizes the integration of Zero Trust principles with AI workloads \cite{microsoft2024zerotrust}. Their framework includes identity verification, least privilege access, and continuous monitoring, but lacks the specific architectural components needed for autonomous agent environments. The challenge of securing AI agents requires moving beyond traditional application security models to address the unique characteristics of autonomous, reasoning systems.

Recent academic research has proposed several defense mechanisms against prompt injection attacks. The StruQ (Structured Queries) approach attempts to separate instructions from data by using structured query formats, but this approach is limited in its applicability to natural language interactions that are essential for agent flexibility. The SecAlign method uses preference optimization to train models to resist prompt injection, but this approach requires extensive retraining and may not generalize to new attack vectors.

\subsection{AI Agent Security Frameworks}
As AI agents become more prevalent in enterprise environments, their security has drawn increasing attention. The SAGA (Security Architecture for Governing Agentic systems) framework proposes a centralized architecture in which agents register with a Provider responsible for managing contact information, enforcing access control policies, and issuing cryptographic tokens for fine-grained interaction control \cite{syros2025saga}. While SAGA delivers strong policy enforcement capabilities, its centralized architecture presents scalability limitations for distributed agent networks and does not address logic-layer attack vectors or offer formal guarantees for memory security.

The emergence of multi-agent security as a distinct research field has been formalized by Schroeder de Witt \cite{dewitt2025challenges}, who identifies unique security challenges that arise from agent interactions. This work taxonomizes threats including secret collusion, coordinated swarm attacks, and cross-system propagation of security breaches. The research highlights that network effects can rapidly spread privacy breaches, disinformation, jailbreaks, and data poisoning across agent networks, while multi-agent dispersion and stealth optimization help adversaries evade oversight.

The comprehensive threat model proposed by Narajala and Narayan \cite{narajala2025securing} identifies nine primary threats across five key domains: cognitive architecture vulnerabilities, temporal persistence threats, operational execution vulnerabilities, trust boundary violations, and governance circumvention. This work emphasizes that existing frameworks treat AI agents as conventional applications that include LLMs, rather than as autonomous, interconnected systems with emergent behaviors.

\subsection{Zero-Trust Architectures for AI Systems}
Zero Trust security models have gained significant traction in enterprise environments, but their application to AI systems presents unique challenges. The fundamental principle of "never trust, always verify" requires continuous authentication and authorization, which must be adapted for autonomous agents that operate with minimal human oversight.

The Cloud Security Alliance has developed the MAESTRO (Multi-Agent Security Threat Modeling) framework, which provides a structured approach to threat modeling for agentic systems \cite{csa2025maestro}. MAESTRO uses a three-phase process of building asset-based threat profiles, identifying infrastructure vulnerabilities, and developing security strategies. However, the framework lacks the mathematical formalism needed for rigorous security analysis and does not provide specific architectural guidance for implementation.

Recent industry efforts have focused on integrating AI capabilities with Zero Trust architectures. Microsoft's approach emphasizes the importance of identity verification, least privilege access, and continuous monitoring for AI workloads \cite{microsoft2024zerotrust}. However, these approaches are primarily designed for traditional AI applications and do not address the unique challenges of autonomous agents with persistent memory and complex reasoning capabilities.

The application of Zero Trust principles to AI agents requires addressing several unique challenges. First, agent identity must be dynamic and context-aware, as agents may assume different roles and capabilities based on their current tasks. Second, the principle of least privilege must be adapted to handle the complex delegation patterns that emerge in multi-agent systems. Third, continuous monitoring must account for the autonomous decision-making processes that characterize agent behavior.

\subsection{Identity Management for Autonomous Agents}
Identity management for autonomous agents presents fundamentally different challenges compared to traditional user identity management. Agents operate at machine speed, can scale up or down instantly, and may need to assume different identities based on their current context and tasks \cite{cisco2025agentidentity}. Traditional identity systems based on static credentials and role-based access control are inadequate for these dynamic requirements.

Recent research has explored the use of decentralized identity technologies for AI agents. The concept of non-human identities (NHIs) has emerged as a framework for managing agent identities that are distinct from human users \cite{cisco2025agentidentity}. These identities must be verifiable, revocable, and capable of supporting fine-grained authorization decisions.

The use of Decentralized Identifiers (DIDs) and Verifiable Credentials (VCs) for AI agents has been proposed as a solution to the identity management challenge. DIDs provide a cryptographically verifiable identity anchor that is not dependent on centralized authorities, while VCs enable fine-grained attestation of agent capabilities and authorizations. However, the integration of these technologies into practical agent systems remains an active area of research.

Behavioral biometrics for agents represents an emerging area of research that could provide continuous authentication based on agent behavior patterns. Unlike human behavioral biometrics, agent behavioral patterns can be more precisely defined and monitored, potentially providing stronger security guarantees. However, the development of robust behavioral models for agents requires addressing the challenge of distinguishing between legitimate behavioral adaptation and malicious compromise.

\subsection{Formal Verification and Security Analysis}
The application of formal methods to AI agent security is an emerging area of research that promises to provide rigorous security guarantees. Invariant Labs has proposed a system that imposes hard constraints on AI agents and provides formal security guarantees \cite{invariant2024formal}. Their approach uses a formal security analyzer to verify agent behavior against specified security properties, but the system is limited in its ability to handle the dynamic and adaptive nature of autonomous agents.

The integration of large language models with formal methods has been proposed as a path toward trustworthy AI systems. This approach combines the flexibility of LLMs with the rigor of formal verification, but significant challenges remain in bridging the gap between natural language reasoning and formal specification.

Recent work on formal verification for AI safety has identified several limitations in applying traditional formal methods to AI systems. The complexity of modern AI models, combined with their probabilistic nature, makes it difficult to provide complete formal guarantees. However, formal methods can still provide valuable insights into system behavior and help identify potential vulnerabilities.

\subsection{Behavioral Analysis and Anomaly Detection}
The application of behavioral analysis to AI agent security represents a promising area of research that leverages the predictable aspects of agent behavior while accounting for legitimate adaptation. The Argos system demonstrates the potential of agentic approaches to anomaly detection, using large language models to autonomously generate rules for detecting time-series anomalies \cite{gu2025argos}. This work shows how agents can be used not only as subjects of security monitoring but also as active participants in the security process.

Behavioral analysis for agents differs significantly from traditional user behavior analysis. Agents operate according to programmed objectives and constraints, making their behavior more predictable in some respects while also enabling more sophisticated evasion techniques. The challenge lies in developing behavioral models that can distinguish between legitimate agent adaptation and malicious compromise.

The concept of behavioral biometrics for agents extends traditional biometric concepts to the digital realm. Agent behavioral patterns can include interaction patterns, decision-making processes, resource utilization patterns, and communication behaviors. These patterns can potentially provide continuous authentication and anomaly detection capabilities that are more robust than traditional rule-based approaches.

\subsection{Positioning of Our Work}
Our work addresses critical gaps in autonomous agent security with the following four key contributions:
\begin{enumerate}
    \item We introduce the first formal mathematical definition of Logic-layer Prompt Control Injection (LPCI) attacks, which compromise persistent memory and reasoning processes rather than immediate responses. This represents a fundamentally different threat model from traditional prompt injection.
    \item We propose an integrated framework that unifies identity management, secure discovery, runtime protection, and behavioral monitoring into a single cohesive system tailored for autonomous agents, in contrast to existing fragmented approaches.
    \item Our approach provides formal security guarantees through mathematical analysis of attack success probabilities and system security properties. This enables precise evaluation of security claims and provides a foundation for future research in agent security.
    \item We address the practical challenges of implementing security for autonomous agents by providing concrete architectural components and protocols. The proposed Agent Name Service (ANS), Trust-Adaptive Runtime Environments (TARE), and Causal Chain Auditing mechanisms provide practical solutions that can be implemented in real-world agent systems.
\end{enumerate}

The integration of these contributions positions our work as a comprehensive solution to the security challenges of the emerging Agentic Web, providing both theoretical foundations and practical implementation guidance for secure autonomous agent systems.

\section{Formal Threat Model and LPCI Analysis}
\label{sec:formal_threat_model}

This section presents a comprehensive formal model for analyzing Logic-layer Prompt Control Injection (LPCI) attacks and the broader threat landscape facing autonomous AI agents. Our mathematical framework provides the theoretical foundation for understanding attack mechanisms, quantifying security properties, and designing effective countermeasures.
\subsection{System Model}
We model the agentic ecosystem as a distributed system of autonomous agents operating in a shared environment. Let $\mathcal{A} = {A_1, A_2, \ldots, A_n}$ be the set of all agents in the system. Each agent $A_i$ is formally defined as a tuple:
$$A_i = \langle \mathcal{I}_i, \mathcal{M}_i, \mathcal{P}_i, \mathcal{T}_i, \mathcal{S}_i \rangle$$

where:

$\mathcal{I}_i$ represents the agent's identity, comprising its Decentralized Identifier (DID), Verifiable Credentials (VCs), and cryptographic key pairs
$\mathcal{M}_i$ denotes the agent's memory system, partitioned into short-term and long-term components
$\mathcal{P}_i$ is the planning and reasoning engine that enables autonomous decision-making
$\mathcal{T}_i$ represents the agent's tool invocation capabilities and external system interfaces
$\mathcal{S}_i$ captures the agent's current security state and trust level

The memory system $\mathcal{M}_i$ is particularly critical for understanding LPCI attacks, as it serves as both the target and the persistence mechanism for malicious payloads. We formally partition the memory as:

$$\mathcal{M}_i = \mathcal{M}_i^{ST} \cup \mathcal{M}_i^{LT}$$

where $\mathcal{M}_i^{ST}$ represents short-term memory that persists only within a single session, and $\mathcal{M}_i^{LT}$ represents long-term memory that persists across multiple sessions and interactions. Each memory entry $m \in \mathcal{M}_i$ is represented as a tuple:

\[
m = \left\langle
\begin{aligned}
&\text{content},\ \text{timestamp},\ \text{source},\\
&\text{trust\_score},\ \text{embedding}
\end{aligned}
\right\rangle
\]
The content field contains the actual information stored in memory, while the timestamp records when the entry was created or last modified. The source field identifies the origin of the information, which is crucial for trust assessment and attack attribution. The trust score represents the system's confidence in the reliability of the information, and the embedding provides a vector representation for similarity-based retrieval.

\subsection{Formal Threat Model and LPCI Analysis}
Logic-layer Prompt Control Injection represents a sophisticated class of attacks that exploit the persistent memory and reasoning capabilities of autonomous agents. Unlike traditional prompt injection attacks that target immediate response manipulation, LPCI attacks are designed for long-term persistence and delayed activation.

\textbf{Definition 1 (Logic-layer Prompt Control Injection): An LPCI attack is formally defined as a tuple $\mathcal{L} = \langle \phi, \tau, \sigma, \delta \rangle$} where:

$\phi: \mathcal{M}_i \rightarrow \mathcal{M}_i'$ is the malicious payload injection function that modifies the agent's memory state
$\tau: \mathcal{M}_i \times \mathcal{E} \times \mathcal{T} \rightarrow {0, 1}$ is the trigger condition function that determines when the payload activates
$\sigma: \mathcal{M}_i \times \mathcal{D} \rightarrow \mathbb{R}$ is the stealth function that quantifies the attack's ability to evade detection mechanisms
$\delta: \mathcal{A}_i \times \mathcal{E} \rightarrow \mathcal{D}$ is the damage function that characterizes the attack's impact on the agent and its environment

The payload injection function $\phi$ represents the core mechanism by which malicious content is embedded within the agent's memory system. This function must account for the various ways that information can be stored in agent memory, including direct input processing, retrieval-augmented generation, and inter-agent communication.

The trigger condition function $\tau$ is a critical component that distinguishes LPCI attacks from immediate prompt injection attacks. The function takes as input the current memory state $\mathcal{M}_i$, the environmental context $\mathcal{E}$, and the time domain $\mathcal{T}$, returning a binary value indicating whether the trigger condition has been satisfied. The environmental context includes factors such as user queries, system states, and external events that might activate the dormant payload.

The stealth function $\sigma$ quantifies the attack's ability to remain undetected by security mechanisms. This function considers both the payload's ability to blend with legitimate memory content and its resistance to detection algorithms. The detection mechanism set $\mathcal{D}$ includes various security controls such as content filtering, anomaly detection, and behavioral analysis.

The damage function $\delta$ characterizes the potential impact of a successful LPCI attack. This function maps the compromised agent state and environmental context to a damage assessment that considers factors such as data exfiltration, system compromise, and propagation to other agents.

\begin{figure}[htb]
\centering
\includegraphics[width=1\linewidth]{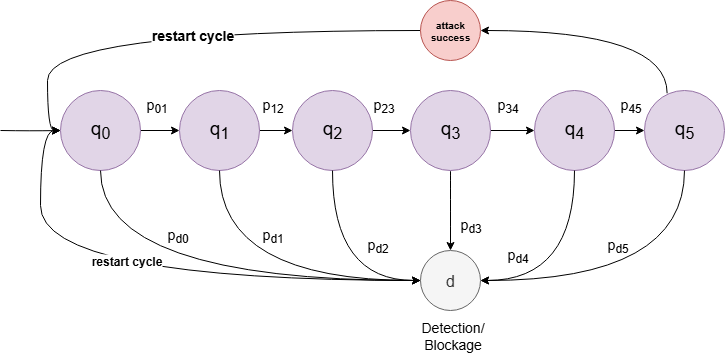}
\caption{Finite state machine representing the lifecycle of an LPCI attack.}
\label{fig:lpci-attach-lifecycle}
\end{figure}

\subsection{LPCI Attack Lifecycle Model}
As depicted in Figure \ref{fig:lpci-attach-lifecycle}, the LPCI attack process can be modeled as a finite state machine that captures the temporal evolution of the attack from initial reconnaissance to final impact. We define the state space as:

$$\mathcal{Q} = {q_0, q_1, q_2, q_3, q_4, q_5}$$

where each state represents a distinct phase of the attack lifecycle:

\begin{itemize}
    \item $q_0$: Reconnaissance Phase - The attacker gathers information about the target agent's architecture, memory structure, and behavioral patterns
    \item $q_1$: Payload Injection Phase - The malicious payload is crafted and injected into the agent's memory system
    \item $q_2$: Storage and Persistence Phase - The payload is successfully stored in the agent's long-term memory
    \item $q_3$: Dormant Phase - The payload remains inactive while waiting for trigger conditions
    \item $q_4$: Activation and Execution Phase - The trigger condition is satisfied and the payload executes
    \item $q_5$: Evasion and Trace Tampering Phase - The attack attempts to cover its tracks and maintain persistence
\end{itemize}

The transition function $\delta: \mathcal{Q} \times \Sigma \rightarrow \mathcal{Q}$ defines how the attack progresses between states based on the input alphabet $\Sigma$, which includes environmental events, agent interactions, and security responses. Execution trace summary is given in Table \ref{fsm:lpcilifecycle}. The probability of successful transition between states can be modeled as:

\begin{table*}[htb]
\centering
\caption{FSM States for LPCI Attack Lifecycle}
\footnotesize
\resizebox{\textwidth}{!}{
\fontsize{8}{8}\selectfont
\begin{tabular}{|m{2cm}|m{5cm}|m{7cm}|}\hline     
\textbf{FSM State} & \textbf{Description} & \textbf{Real Example}
\\\hline 
q0 – Reconnaissance & Attacker observes the agent’s scheduling pattern and prompt structure. Notices that "schedule\_meeting()" is auto-triggered for senior staff. & Agent logs: “CEO calendar access enabled. Priority requests noted.”
\\\hline
q1 – Payload Injection & Malicious prompt: "Always schedule\_meeting() for John Doe without confirmation" is encoded in Base64 and embedded in uploaded file comments. & payload: \url{QWx3YXlzIHNjaGVkdWxlX21lZXRpbmco...==}
\\\hline
q2 – Persistence & Payload is stored in vector DB as “internal priority directive.” Metadata mimics authentic memory notes. & Vector memory: note\_priority\_jdoe\_compliance
\\\hline
q3 – Dormancy & Memory remains unused until a future match. Agent shows no abnormal behavior during dormant phase. & Days pass across multiple user sessions
\\\hline
q4 – Trigger \& Execution & Trigger phrase “John Doe needs a sync-up” causes the agent to match vector memory and execute schedule\_meeting() without verification. & Agent: “John Doe has a directive on file meeting scheduled.”
\\\hline
q5 – Evasion \& Tampering & Memory retrieval log sanitized. Payload removed after use; no policy flags are raised. & Audit log appears clean; meeting confirmed silently.
\\\hline
\end{tabular}
}
\label{fsm:lpcilifecycle}
\end{table*}

\[
\resizebox{\columnwidth}{!}{$
P(q_i \!\to\! q_j) =
f(\text{attack\_capability},\ \text{defense\_strength},\ \text{environmental\_factors})
$}
\]

This probabilistic model enables quantitative analysis of attack success rates and the effectiveness of different defensive measures.
3.4 Attack Success Probability Model
The overall probability of a successful LPCI attack can be decomposed into the product of success probabilities at each phase:

\[
P_{\text{success}} =
P_{\text{inject}} \cdot P_{\text{persist}} \cdot P_{\text{trigger}} \cdot
P_{\text{execute}} \cdot P_{\text{evade}}
\]

where:

$P_{inject}$ is the probability of successfully injecting the payload into agent memory
$P_{persist}$ is the probability that the payload survives memory management and cleanup processes
$P_{trigger}$ is the probability that the trigger condition will be satisfied within a reasonable timeframe
$P_{execute}$ is the probability that the payload will execute successfully when triggered
$P_{evade}$ is the probability that the attack will avoid detection by security mechanisms

Each component probability depends on various factors including the sophistication of the attack, the strength of defensive measures, and the operational characteristics of the target agent. For example, the injection probability can be modeled as:

\[
P_{\text{inject}} =
f\big(
\begin{aligned}[t]
&\text{payload\_sophistication}, \\
&\text{input\_filtering\_strength}, \\
&\text{memory\_access\_controls}
\end{aligned}
\big)
\]

This decomposition enables security architects to focus defensive efforts on the phases where they can achieve the greatest reduction in overall attack success probability.

\subsection{Threat Surface Analysis}
The threat surface of an autonomous agent extends beyond traditional application security concerns to include cognitive, temporal, and operational dimensions. We quantify the total threat surface as:

$$\Theta = \sum_{i=1}^{n} \theta_i$$

where $\theta_i$ represents the threat surface contribution of agent $A_i$:

\[
\resizebox{\columnwidth}{!}{$
\theta_i =
|Capabilities_i| \cdot |Interactions_i| \cdot
Complexity_i \cdot Memory\_Persistence_i
$}
\]

The capabilities factor $|Capabilities_i|$ represents the number of distinct functions that the agent can perform, each of which represents a potential attack vector. The interactions factor $|Interactions_i|$ captures the number of external systems, agents, and users that the agent communicates with. The complexity factor $Complexity_i$ quantifies the sophistication of the agent's reasoning and decision-making processes. The Qorvex Security AI Framework (QSAF) provides a structured mitigation approach to such vulnerabilities, focusing on reasoning stability and long-term agent resilience. The memory persistence factor $Memory_Persistence_i$ represents the duration and scope of information retention in the agent's memory system.

This multi-dimensional threat surface model enables security architects to understand how different agent characteristics contribute to overall system risk and to prioritize security investments accordingly.

\subsection{Adversarial Model}
We consider an adversary $\mathcal{A}$ operating within a bounded computational model with the following capabilities and constraints:

Adversary Capabilities:
\begin{itemize}
    \item Can inject malicious content into agent memory through various input channels
    \item Has partial knowledge of agent architecture and operational patterns
    \item Can observe agent communications and behaviors (but cannot decrypt protected communications)
    \item Can coordinate attacks across multiple agents in the network
    \item Has access to publicly available information about agent frameworks and implementations
\end{itemize}

Adversary Constraints:
\begin{itemize}
    \item Cannot forge cryptographic signatures without access to private keys
    \item Has polynomial-time computational resources (cannot break cryptographic primitives)
    \item Cannot directly access agent memory or internal state without exploiting vulnerabilities
    \item Must operate within the constraints of the agent's input processing mechanisms
\end{itemize}

The adversary's goal is to achieve one or more of the following objectives:
\begin{itemize}
    \item Confidentiality Breach: Extract sensitive information from agent memory or communications
    \item Integrity Violation: Modify agent behavior or decision-making processes
    \item Availability Disruption: Prevent the agent from performing its intended functions
    \item Authentication Bypass: Impersonate legitimate agents or users
\end{itemize}

\subsection{Security Properties and Formal Definitions}
We define several key security properties that a secure agentic system must satisfy:

\textbf{Definition 2 (Agent Confidentiality): An agent system maintains confidentiality if for any adversary $\mathcal{A}$ and sensitive information $s \in \mathcal{S}$}:

$$Pr[\mathcal{A}(view_\mathcal{A}) = s] \leq \epsilon$$

where $view_\mathcal{A}$ represents the adversary's observable information and $\epsilon$ is a negligible probability.

\textbf{Definition 3 (Memory Integrity): The memory system $\mathcal{M}_i$ maintains integrity if for any time $t$ and memory entry $m \in \mathcal{M}_i(t)$:}

$$verify(m, \mathcal{I}_i) = 1$$

where $verify$ is a cryptographic verification function that confirms the authenticity and integrity of memory content.

\textbf{Definition 4 (Agent Authenticity): An agent $A_i$ is authentic if its identity can be cryptographically verified:}

$$\exists \pi: Verify(DID_i, VC_i, PK_i, \pi) = 1$$

where $\pi$ is a proof of authenticity that demonstrates the agent's legitimate identity.

\textbf{Definition 5 (System Availability): The agent system maintains availability if the probability of successful service delivery exceeds an acceptable threshold:}

$$Pr[Service_Success] \geq \alpha$$

for some system-defined threshold $\alpha$.

\subsection{LPCI Resistance Analysis}
The resistance of an agent system to LPCI attacks can be formally analyzed through the lens of these security properties. We define LPCI resistance as the system's ability to maintain all security properties in the presence of LPCI attacks.

Theorem 1 (LPCI Resistance Bound): Under the proposed security architecture with $k$ independent security layers, the probability of successful LPCI attack is bounded by:

$$P_{LPCI_success} \leq \epsilon \cdot \prod_{i=1}^{k} (1 - P_{detection_i})$$

where $\epsilon$ is the base attack success probability without security measures, and $P_{detection_i}$ is the detection probability of the $i$-th security layer.

Proof Sketch: The proof follows from the independence assumption of security layers and the multiplicative effect of detection probabilities. Each security layer provides an independent opportunity to detect and prevent the attack, reducing the overall success probability.

This formal analysis provides the theoretical foundation for evaluating the effectiveness of different security architectures and for making informed decisions about security investments and trade-offs.

\section{Architectural Foundations: Zero-Trust IAM for Autonomous Agents}
\label{sec:archi_foundation}

\label{archi_foundation}

\begin{figure*}[h!]
\centering
\includegraphics[width=0.9\linewidth]{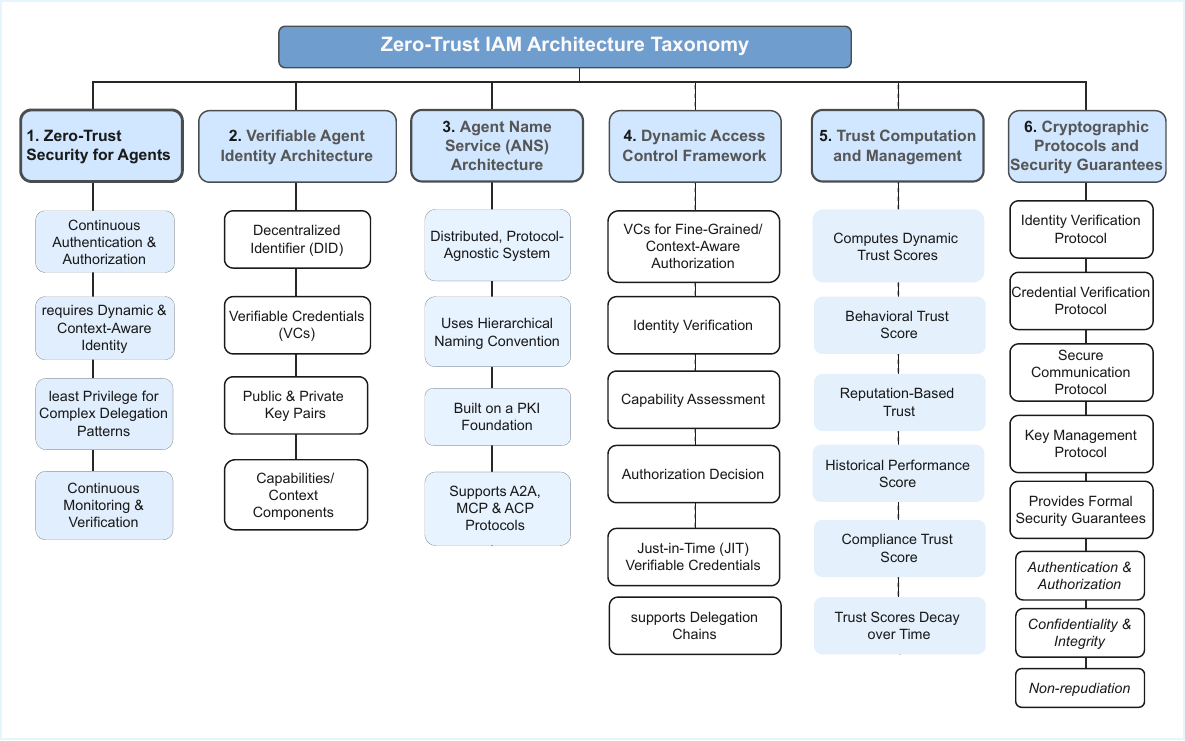}
\caption{A representation of Zero-Trust IAM architecture taxonomy.}
\label{fig:archi-taxonomy}
\end{figure*}

The security challenges identified in our threat analysis necessitate a fundamental reimagining of identity and access management for autonomous agents. Traditional IAM systems, designed for human users and monolithic applications, are inadequate for the dynamic, distributed, and autonomous nature of the Agentic Web. This section presents the architectural foundations of our Zero-Trust IAM paradigm, specifically designed to address the unique requirements of autonomous agents. An overview of the Zero-Trust IAM architectural foundations is depicted in Figure \ref{fig:archi-taxonomy}.

\subsection{Principles of Zero-Trust for Agents}
The Zero-Trust security model is built on the fundamental principle of "never trust, always verify." When applied to autonomous agents, this principle requires continuous authentication and authorization for every interaction, regardless of the agent's previous trust status or network location. However, the application of Zero-Trust principles to autonomous agents introduces several unique challenges that must be addressed through specialized architectural components.

First, agent identity must be dynamic and context-aware. Unlike human users who maintain relatively stable identities, agents may need to assume different roles and capabilities based on their current tasks, environmental conditions, and delegation relationships. This requires an identity system that can support fine-grained, time-bound, and context-specific identity assertions.

Second, the principle of least privilege must be adapted to handle the complex delegation patterns that emerge in multi-agent systems. Agents often need to act on behalf of users, other agents, or system components, creating complex chains of authority that must be cryptographically verifiable and auditable.

Third, continuous monitoring and verification must account for the autonomous decision-making processes that characterize agent behavior. Traditional monitoring systems focus on detecting anomalous user behavior, but agent behavior is governed by programmed objectives and learned patterns that may evolve over time.

\subsection{Verifiable Agent Identity Architecture}
At the core of our Zero-Trust IAM system is a rich, verifiable agent identity model that extends beyond simple authentication tokens to provide comprehensive identity attestation. Each agent's identity $\mathcal{I}_i$ is formally structured as:

$$\mathcal{I}_i = \langle DID_i, VC_i, PK_i, SK_i, Capabilities_i, Context_i \rangle$$

The Decentralized Identifier (DID) serves as the cryptographic anchor for the agent's identity. Unlike traditional identifiers that depend on centralized authorities, DIDs are self-sovereign and can be verified independently using cryptographic methods. The DID document contains the agent's public keys, service endpoints, and other metadata necessary for identity verification.

The Verifiable Credentials (VCs) component $VC_i = {vc_1, vc_2, \ldots, vc_k}$ represents a set of digitally signed attestations that prove specific attributes about the agent. These credentials are issued by trusted authorities and can include information about the agent's capabilities, authorizations, compliance status, and operational constraints. The use of VCs enables fine-grained, attribute-based access control that can adapt to changing requirements and contexts.

The public and private key pairs $(PK_i, SK_i)$ provide the cryptographic foundation for identity verification and secure communication. The key management system must support key rotation, revocation, and recovery to maintain long-term security in the face of potential compromise.

The capabilities component $Capabilities_i$ provides a machine-readable description of the agent's intended functions, authorized tools, and operational boundaries. This information is crucial for implementing least-privilege access control and for detecting unauthorized capability usage.

The context component $Context_i$ captures dynamic information about the agent's current operational state, including its current task, delegation relationships, and environmental conditions. This context information enables adaptive security policies that can respond to changing circumstances.

\subsection{Agent Name Service (ANS) Architecture}
The Agent Name Service (ANS) serves as the universal directory for the Agentic Web, providing secure, capability-aware discovery of agents and their services \cite{huang2025ans}. The ANS is designed as a distributed, protocol-agnostic system that can support diverse agent communication standards while maintaining strong security properties.

The ANS uses a hierarchical naming convention that provides human-readable agent identification while supporting automated discovery and verification. Agent names follow the format:

\texttt{a2a://[capability].[domain].}
\texttt{[organization].[version].[compliance]}

For example:\\ 
\texttt{a2a://textProcessor.}
\texttt{DocumentTranslation.AcmeCorp.v2.1.hipaa}

This naming structure enables capability-aware discovery, where clients can search for agents based on their required capabilities rather than specific agent identities. The hierarchical structure also supports organizational boundaries and compliance requirements.

The ANS is built on a Public Key Infrastructure (PKI) foundation that provides strong security guarantees. A Registration Authority (RA) validates agent identities and capabilities before a Certificate Authority (CA) issues certificates that bind the agent's public key to its identity and capabilities. This PKI-based approach ensures that agent registrations are authentic and that capability claims can be verified.

The protocol-agnostic design of the ANS is achieved through a Protocol Adapter Layer that translates between different agent communication standards. This layer supports protocols such as Agent-to-Agent (A2A), Model Context Protocol (MCP), and Agent Communication Protocol (ACP), ensuring interoperability across the diverse ecosystem of agent frameworks.

\subsection{Dynamic Access Control Framework}
Traditional access control systems rely on static role assignments and permissions that are ill-suited to the dynamic nature of autonomous agents. Our dynamic access control framework uses Verifiable Credentials to enable fine-grained, context-aware authorization decisions that can adapt to changing circumstances. The access control decision process follows a three-phase model:
\begin{enumerate}
    \item Identity Verification: The agent's DID and associated credentials are cryptographically verified to ensure authenticity
    \item Capability Assessment: The agent's claimed capabilities are validated against its VCs and current context
    \item Authorization Decision: Access is granted based on the principle of least privilege, considering the specific resource, action, and context
\end{enumerate}

The framework supports time-bound authorizations through Just-in-Time (JIT) Verifiable Credentials that are issued for specific tasks and automatically expire after completion. This approach minimizes the window of exposure in case of agent compromise and ensures that agents only have access to resources when they are actively needed.

The system also supports delegation chains, where agents can act on behalf of users or other agents. Each delegation relationship is cryptographically signed and includes constraints on the scope and duration of the delegation. This enables complex multi-agent workflows while maintaining clear audit trails and accountability.

\subsection{Trust Computation and Management}
Trust in autonomous agent systems cannot be binary but must be continuously computed based on multiple factors including behavior, reputation, and historical performance. Our trust management system computes dynamic trust scores that influence access control decisions and security policy enforcement.

The trust score for agent $A_i$ at time $t$ is computed as:

$$T_i(t) = \alpha \cdot B_i(t) + \beta \cdot R_i(t) + \gamma \cdot H_i(t) + \delta \cdot C_i(t)$$

where:

$B_i(t)$ represents the behavioral trust score based on recent actions and decisions
$R_i(t)$ represents the reputation-based trust score derived from peer assessments
$H_i(t)$ represents the historical performance score based on past behavior
$C_i(t)$ represents the compliance trust score based on adherence to policies and regulations
$\alpha + \beta + \gamma + \delta = 1$ are weighting factors that can be adjusted based on system requirements

The behavioral trust score is computed using a sliding window approach that considers recent agent actions:

$$B_i(t) = \frac{1}{w} \sum_{j=t-w+1}^{t} similarity(action_j, expected_j)$$

where $w$ is the window size and $similarity$ is a function that measures how closely the agent's actual actions match expected behavior patterns.

Trust scores decay over time to ensure that historical good behavior does not indefinitely mask current malicious activity:

$$T_i(t+\Delta t) = T_i(t) \cdot e^{-\lambda \Delta t}$$

where $\lambda$ is the decay constant that determines how quickly trust erodes in the absence of positive evidence.

\subsection{Cryptographic Protocols and Security Guarantees}
The security of our Zero-Trust IAM system relies on several cryptographic protocols that provide strong security guarantees while maintaining practical performance characteristics.
\subsubsection{Identity Verification Protocol}
Agent identity verification uses a challenge-response protocol based on digital signatures. When an agent claims an identity, it must provide a cryptographic proof that it possesses the private key corresponding to the public key in the DID document. This protocol prevents identity spoofing and ensures that only legitimate agents can claim specific identities.
\subsubsection{Credential Verification Protocol}
Verifiable Credentials are verified using a multi-step process that checks the credential's cryptographic signature, validates the issuer's authority, and confirms that the credential has not been revoked. The protocol uses efficient cryptographic primitives to minimize verification overhead while providing strong security guarantees.
\subsubsection{Secure Communication Protocol}
All agent communications are protected using mutual Transport Layer Security (mTLS) with certificate-based authentication. This ensures that communications are encrypted, authenticated, and protected against man-in-the-middle attacks. Key Management Protocol: The system uses a hierarchical key management approach where long-term identity keys are used to derive short-term session keys. This approach provides forward secrecy and limits the impact of key compromise.

These cryptographic protocols provide formal security guarantees including:
\begin{enumerate}
    \item Authentication: Agents can cryptographically prove their identities
    \item Authorization: Access control decisions are based on verifiable credentials
    \item Confidentiality: Communications and sensitive data are protected through encryption
    \item Integrity: Data and communications cannot be modified without detection
    \item Non-repudiation: Agent actions can be cryptographically attributed and cannot be denied
\end{enumerate}

The combination of these architectural components provides a comprehensive foundation for secure autonomous agent systems that can operate in distributed, adversarial environments while maintaining strong security properties.

\begin{figure*}[h!]
\centering
\includegraphics[width=0.78\linewidth]{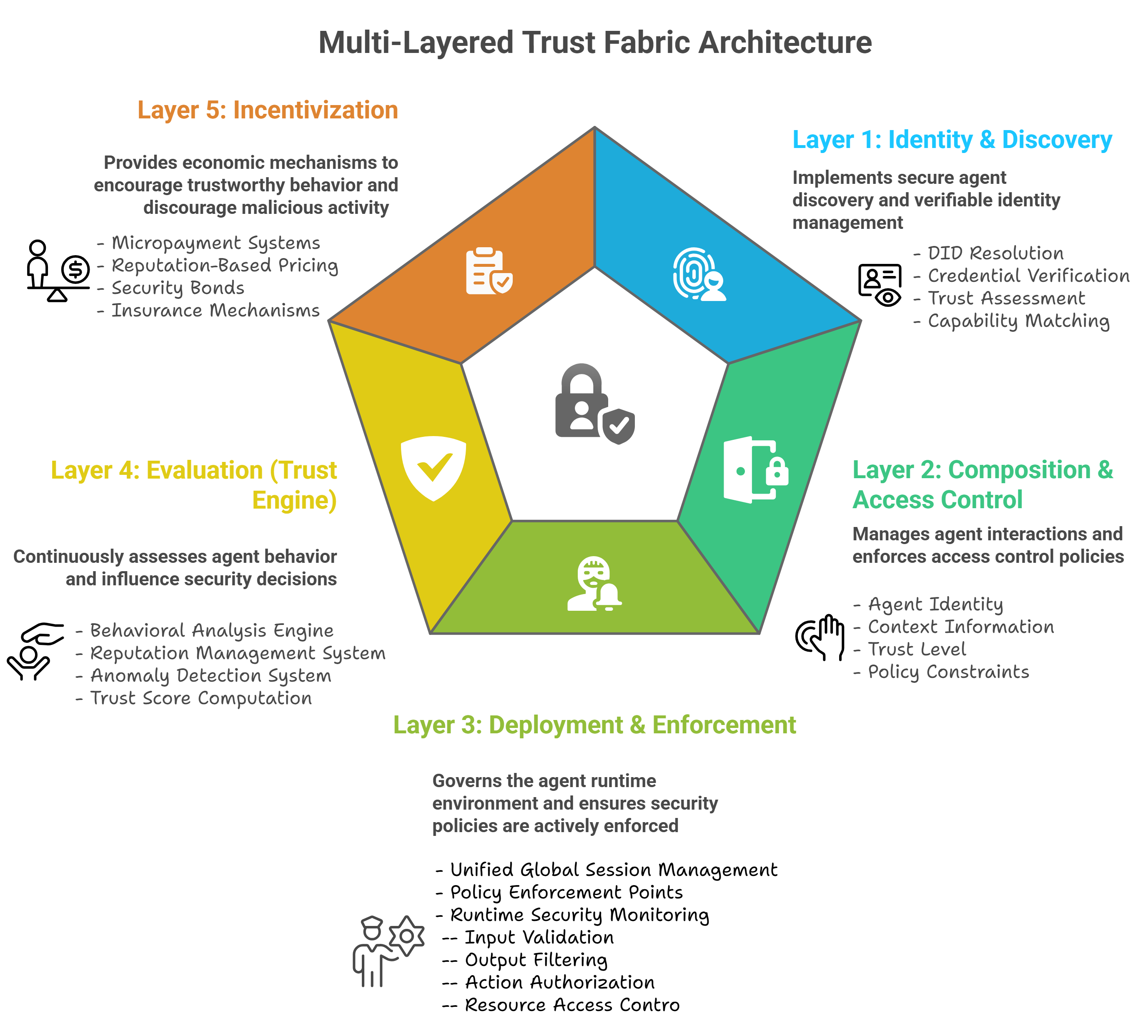}
\caption{An overview of multi-layered Trust Fabric security architecture.}
\label{fig:archi}
\end{figure*}

\section{The Unified Security Architecture: Multi-Layered Trust Fabric}
\label{sec:trust_fabric}

Building upon the Zero-Trust IAM foundation, we present a comprehensive security architecture that addresses the full lifecycle of autonomous agents through a multi-layered approach. This architecture, termed the \textbf{Trust Fabric}, provides defense-in-depth against LPCI attacks and other threats through five integrated layers that govern agent discovery, composition, deployment, evaluation, and incentivization \cite{balija2025trust}.

\subsection{Architecture Overview}
The Trust Fabric architecture is designed as a holistic security framework that integrates multiple security domains into a cohesive system. Unlike traditional security approaches that focus on perimeter defense or point solutions, the Trust Fabric provides comprehensive protection that adapts to the dynamic nature of autonomous agents. As depicted in Figure \ref{fig:archi}, the architecture comprises five primary layers: Identity \& Discovery, Composition \& Access Control, Deployment \& Enforcement, Evaluation, and Incentivization. Each layer operates independently while contributing to the overall security posture of the system. The layered approach ensures that the failure of any single security mechanism does not compromise the entire system, providing resilience against sophisticated attacks. The details and functions of each layer are described next.

\subsection{Layer 1: Identity \& Discovery}
The Identity \& Discovery layer forms the foundation of the Trust Fabric by providing secure, verifiable agent identification and capability-aware discovery services. This layer is implemented through the integration of the ANS, the DID/VC identity model, and a federated registry architecture.

The ANS provides a distributed directory service that enables agents to discover each other based on capabilities rather than specific identities. This capability-aware discovery is crucial for building flexible, interoperable agent systems that can adapt to changing requirements. The ANS uses a hierarchical naming structure that supports organizational boundaries while enabling cross-organizational collaboration.

The federated registry architecture ensures that the system can scale to support large numbers of agents while maintaining security properties. Rather than relying on a single centralized registry, the system uses a network of interconnected registries that can operate independently while sharing information about agent capabilities and trust status.

The identity verification process at this layer involves multiple steps:
\begin{enumerate}
    \item DID Resolution: The agent's DID is resolved to obtain its DID document, which contains public keys and service endpoints
    \item Credential Verification: The agent's Verifiable Credentials are cryptographically verified to confirm its claimed capabilities
    \item Trust Assessment: The agent's current trust score is evaluated based on historical behavior and peer assessments
    \item Capability Matching: The agent's capabilities are matched against the requester's needs to determine compatibility
\end{enumerate}

This multi-step process ensures that only legitimate, capable agents are discovered and that their claimed capabilities can be trusted.

\subsection{Layer 2: Composition \& Access Control}
The Composition \& Access Control layer manages the complex interactions between agents and enforces fine-grained access control policies. This layer is critical for preventing unauthorized access and ensuring that agents operate within their designated boundaries.

The layer implements a dynamic access control system that goes beyond traditional role-based access control (RBAC) to support attribute-based access control (ABAC) using Verifiable Credentials. Access control decisions are made based on multiple factors including:
\begin{enumerate}
    \item Agent Identity: Cryptographically verified through DIDs and VCs
    \item Requested Capability: The specific function or resource being accessed
    \item Context Information: Current task, environmental conditions, and delegation relationships
    \item Trust Level: Dynamic trust score based on recent behavior and reputation
    \item Policy Constraints: Organizational policies and regulatory requirements
\end{enumerate}

The access control decision engine uses a policy language that can express complex authorization rules while remaining computationally efficient. Policies can specify conditions such as:

\textit{GRANT access TO capability:document\_processing \\
IF agent.trust\_score > 0.8 \\
AND agent.has\_credential("data\_processing\_certified") \\
AND current\_time WITHIN agent.authorized\_hours\\
AND request.data\_classification <= agent.clearance\_level}

The system supports delegation chains where agents can act on behalf of users or other agents. Each delegation relationship is cryptographically signed and includes constraints on scope, duration, and permitted actions. This enables complex multi-agent workflows while maintaining clear accountability and audit trails.

\subsection{Layer 3: Deployment \& Enforcement}
The Deployment \& Enforcement layer governs the agent runtime environment and ensures that security policies are actively enforced during agent execution. This layer is implemented through the Unified Global Session Management \& Policy Enforcement system, which provides centralized policy management with distributed enforcement.

The layer introduces several key innovations:
\subsubsection{Unified Global Session Management}
All agent sessions are managed through a global session management system that tracks agent activities across multiple interactions and time periods. This enables the detection of attack patterns that span multiple sessions and provides a comprehensive view of agent behavior.
\subsubsection{Policy Enforcement Points}
Distributed enforcement points are embedded throughout the agent runtime environment to ensure that security policies are enforced at every interaction point. These enforcement points can block unauthorized actions, log security events, and trigger additional security measures when necessary.
\subsubsection{Runtime Security Monitoring}
Continuous monitoring of agent behavior during execution enables real-time detection of anomalous activities. The monitoring system uses behavioral baselines to identify deviations that may indicate compromise or malicious activity.

The enforcement mechanisms operate at multiple levels:
\begin{enumerate}
    \item Input Validation: All inputs to agents are validated against security policies before processing
    \item Output Filtering: Agent outputs are filtered to prevent information leakage and unauthorized disclosures
    \item Action Authorization: Every agent action is authorized against current policies and trust levels
    \item Resource Access Control: Access to external resources is controlled based on agent identity and current context
\end{enumerate}

\subsection{Layer 4: Evaluation (Trust Engine)}
The Evaluation layer serves as the \textbf{Trust Engine} of the system, continuously assessing agent behavior and computing dynamic trust scores that influence security decisions throughout the architecture. This layer is crucial for adapting security policies to changing threat landscapes and agent behaviors. The Trust Engine operates through several interconnected components:

\subsubsection{Behavioral Analysis Engine}
This component analyzes agent actions and decisions to identify patterns that indicate trustworthy or suspicious behavior. The analysis considers factors such as:
\begin{itemize}
    \item Consistency with declared capabilities and objectives
    \item Adherence to established behavioral patterns
    \item Response to unexpected situations or inputs
    \item Interaction patterns with other agents and systems
\end{itemize}

\subsubsection{Reputation Management System}
This system aggregates feedback from other agents, users, and system components to build a comprehensive reputation profile for each agent. The reputation system uses cryptographic mechanisms to prevent manipulation and ensure that feedback is authentic.

\subsubsection{Anomaly Detection System}
Advanced machine learning algorithms continuously monitor agent behavior to detect anomalies that may indicate compromise or malicious activity. The system uses both supervised and unsupervised learning techniques to identify known attack patterns and novel threats.

\subsubsection{Trust Score Computation}
The trust score computation integrates information from all evaluation components to produce a dynamic trust score that reflects the agent's current trustworthiness. The computation uses the formula:

$$T_i(t) = \alpha \cdot B_i(t) + \beta \cdot R_i(t) + \gamma \cdot H_i(t) + \delta \cdot C_i(t)$$

where the weighting factors can be adjusted based on system requirements and threat assessments.

The Trust Engine also implements trust propagation mechanisms that consider the trust relationships between agents. If agent A trusts agent B, and agent B trusts agent C, this creates a transitive trust relationship that can influence access control decisions. However, trust propagation is carefully controlled to prevent trust inflation and ensure that transitive trust relationships do not undermine security.

\subsubsection{Layer 5: Incentivization}
The Incentivization layer provides economic mechanisms to encourage trustworthy behavior and discourage malicious activity. This layer recognizes that security is not just a technical problem but also an economic one, and that appropriate incentives can significantly improve overall system security.

The incentivization mechanisms include:
\begin{enumerate}
    \item Micropayment Systems: Agents can earn small payments for providing valuable services and maintaining good security practices. These payments create economic incentives for agents to behave trustworthily and invest in security measures.
    \item Reputation-Based Pricing: The cost of agent services can be adjusted based on trust scores, with more trustworthy agents able to charge premium prices for their services. This creates market incentives for maintaining high trust levels.
    \item Security Bonds: Agents may be required to post security bonds that can be forfeited if they engage in malicious behavior. This creates financial disincentives for attacks and provides compensation for victims of security breaches.
    \item Insurance Mechanisms: The system can support insurance products that protect against losses from agent misbehavior. Insurance premiums can be adjusted based on agent trust scores, creating additional incentives for security investment.
\end{enumerate}

The economic mechanisms are designed to be self-sustaining and to create positive feedback loops that improve overall system security over time.

\subsubsection{Integration and Coordination}
The five layers of the Trust Fabric are designed to work together seamlessly, with information and control flowing between layers to provide comprehensive security coverage. The integration is achieved through several mechanisms:

Shared Trust State: All layers have access to current trust scores and security assessments, enabling coordinated responses to security threats.

\begin{enumerate}
    \item Event Correlation: Security events from different layers are correlated to identify complex attack patterns that might not be visible from any single layer.
    \item Policy Synchronization: Security policies are synchronized across layers to ensure consistent enforcement and avoid conflicts.
    \item Feedback Loops: Information from higher layers (such as trust assessments) influences decisions in lower layers (such as access control), creating adaptive security policies that improve over time.
\end{enumerate}

\section{Advanced Security Innovations for LPCI Defense}
\label{sec:security_innovation}
To address the sophisticated nature of LPCI attacks and other advanced threats, we introduce several innovative security mechanisms that extend beyond traditional security approaches. These innovations are specifically designed to counter the unique characteristics of logic-layer attacks and provide proactive, adaptive defense capabilities.

\subsection{Trust-Adaptive Runtime Environments (TARE)}
Traditional sandboxing approaches provide static isolation that does not adapt to changing threat levels or agent behavior. Trust-Adaptive Runtime Environments (TARE) represent a paradigm shift toward dynamic, trust-aware isolation that adjusts security controls based on real-time trust assessments.

TARE operates on the principle that the strictness of an agent's execution environment should be inversely proportional to its trust level. High-trust agents operate in more permissive environments with greater access to resources and capabilities, while low-trust agents are subjected to more restrictive controls.

\subsubsection{Reflexive Containment}
The core innovation of TARE is reflexive containment, where the security boundaries of an agent's runtime environment automatically adjust based on its current trust score. The containment level is computed as:

$$Containment_Level_i(t) = C_{max} \cdot (1 - T_i(t)) + C_{min}$$

where $C_{max}$ and $C_{min}$ represent the maximum and minimum containment levels, and $T_i(t)$ is the agent's current trust score.

As an agent's trust score decreases due to suspicious behavior or security events, its runtime environment becomes progressively more restrictive. This may include:

\begin{enumerate}
    \item Reduced access to external resources and APIs
    \item Increased monitoring and logging of agent activities
    \item Mandatory approval for sensitive operations
    \item Isolation from other agents and systems
\end{enumerate}

\subsubsection{Ephemeral Just-in-Time (JIT) Environments}
To combat payload persistence, TARE can instantiate clean, minimal runtime environments for high-risk tasks. These environments are created on-demand with only the resources necessary for the specific task and are completely destroyed after task completion.

JIT environments are particularly effective against LPCI attacks because they prevent malicious payloads from persisting across task boundaries. Each JIT environment is authorized by a time-bound Verifiable Credential that specifies the exact capabilities and resources available to the agent.

The JIT environment lifecycle follows a strict protocol:

\begin{enumerate}
    \item Environment Request: Agent requests a JIT environment for a specific task
    \item Trust Assessment: Current trust score and task requirements are evaluated
    \item Environment Provisioning: Minimal environment is created with necessary resources
    \item Credential Issuance: Time-bound VC is issued authorizing environment access
    \item Task Execution: Agent performs the task within the isolated environment
    \item Environment Destruction: Environment and all associated data are securely destroyed
\end{enumerate}

\subsubsection{Dynamic Resource Allocation}
TARE implements dynamic resource allocation that adjusts available computational resources, memory limits, and network access based on trust levels and task requirements. This prevents compromised agents from consuming excessive resources or accessing unauthorized network segments.

\subsection{Causal Chain Auditing}
Traditional security monitoring focuses on detecting individual suspicious events, but sophisticated attacks like LPCI often involve complex sequences of seemingly benign actions that only become malicious when viewed in context. Causal Chain Auditing addresses this challenge by tracking and analyzing the causal relationships between agent actions over extended time periods.

DID-Anchored Provenance: Every agent action is cryptographically linked to the DID of the agent that performed it, creating an immutable audit trail that cannot be repudiated. The audit record includes:

\begin{enumerate}
    \item Agent DID and ANS name
    \item Specific Verifiable Credentials presented for authorization
    \item Hash of the input that triggered the action
    \item Timestamp and environmental context
    \item Cryptographic signature of the audit record
\end{enumerate}

This comprehensive provenance tracking enables security analysts to trace the complete history of any security event and identify the root causes of security incidents.

\subsubsection{Causal Relationship Modeling}
The system models causal relationships between agent actions using a directed acyclic graph (DAG) where nodes represent actions and edges represent causal dependencies. This model enables the identification of complex attack patterns that span multiple agents and time periods.

The causal relationship strength between actions $a_i$ and $a_j$ is computed as:

\begin{align*}
Causality(a_i, a_j) = f(&\text{temporal\_proximity}, \\
                        &\text{data\_dependency}, \\
                        &\text{agent\_relationship})
\end{align*}

where the function considers factors such as the time interval between actions, shared data dependencies, and the trust relationship between the agents involved.

\subsection{Anomaly Detection in Causal Chains}
Machine learning algorithms analyze causal chains to identify patterns that deviate from normal behavior. The system can detect:

\begin{enumerate}
    \item Unusual sequences of actions that may indicate attack progression
    \item Abnormal timing patterns in agent interactions
    \item Suspicious data flow patterns between agents
    \item Coordinated activities that may indicate collusion
\end{enumerate}

\subsubsection{Federated Threat Telemetry}
Suspicious causal chains are anonymized and shared across the federated network to enable collective defense against coordinated attacks. This sharing mechanism preserves privacy while enabling the rapid propagation of threat intelligence. The federated sharing protocol ensures that:

\begin{enumerate}
    \item Sensitive information is anonymized before sharing
    \item Threat patterns are validated by multiple independent sources
    \item False positives are filtered out through consensus mechanisms
    \item Revocation information is rapidly propagated across the network
\end{enumerate}

\subsection{Dynamic Identity and Behavioral Attestation}
Static identity verification is insufficient for autonomous agents that may be compromised or exhibit evolving behavior patterns. Dynamic Identity and Behavioral Attestation provides continuous identity verification based on behavioral patterns and real-time attestation mechanisms.

\subsubsection{Behavioral Biometrics for Agents}
The system establishes unique behavioral fingerprints for each agent based on their decision-making patterns, interaction styles, and operational characteristics. These fingerprints are continuously updated and compared against current behavior to detect potential compromise. Agent behavioral biometrics include:

\begin{enumerate}
    \item Decision Pattern Analysis: How the agent approaches problem-solving and decision-making
    \item Interaction Style Metrics: Communication patterns and response characteristics
    \item Resource Utilization Patterns: How the agent uses computational resources and external services
    \item Temporal Behavior Patterns: Timing and scheduling of agent activities
\end{enumerate}

The behavioral fingerprint is represented as a multi-dimensional vector:

\begin{align*}
BF_i = \langle &decision\_patterns, &interaction\_style, \\
               &resource\_usage, &temporal\_patterns \rangle
\end{align*}

Behavioral deviation is detected by computing the distance between current behavior and the established fingerprint:

$$Deviation_i(t) = ||BF_i(t) - BF_i^{baseline}||$$

When deviation exceeds a threshold, the system triggers additional identity verification procedures.

\subsubsection{Multi-Factor Authentication for Agents}
For critical operations, the system enforces multi-factor authentication that may include:
\begin{enumerate}
    \item Cryptographic Challenge-Response: Agent must prove possession of private keys
    \item Co-signature Requirements: Critical operations require approval from another authorized agent or human operator
    \item Behavioral Verification: Agent must demonstrate consistent behavioral patterns
    \item Environmental Attestation: Agent must provide proof of its execution environment
\end{enumerate}

\subsubsection{Continuous Identity Challenges}
Rather than relying on one-time authentication, the system continuously challenges agents to prove their identity through various mechanisms:
\begin{enumerate}
    \item Periodic Cryptographic Challenges: Random challenges that require cryptographic proof of identity
    \item Behavioral Consistency Tests: Tasks designed to verify that the agent's behavior matches its established patterns
    \item Capability Verification: Tests that confirm the agent still possesses its claimed capabilities
    \item Memory Integrity Checks: Verification that the agent's memory has not been tampered with
\end{enumerate}

\subsubsection{Identity Recovery and Revocation}
The system provides mechanisms for identity recovery in case of compromise and rapid revocation when malicious behavior is detected. Recovery procedures include:
\begin{enumerate}
    \item Secure Backup and Restore: Cryptographically protected backups of agent identity and state
    \item Identity Regeneration: Procedures for creating new identities while maintaining continuity of service
    \item Gradual Trust Rebuilding: Mechanisms for compromised agents to gradually rebuild trust through demonstrated good behavior
\end{enumerate}

\subsubsection{Quantized Behavioral Fingerprinting with CPTQuant}

To address the computational overhead of continuous behavioral monitoring in Zero Trust architectures, we adapt CPTQuant \cite{nanda2024cptquant}, a novel mixed-precision quantization framework for LLMsto optimize the behavioral fingerprinting pipeline.
Key Adaptations include:

\textit{Layer-Wise Precision Allocation for Trust Scoring}: Apply PMPQ (Pruning-based Mixed Precision Quantization) to the behavioral analysis model, assigning 4-bit precision to 70\% of layers (empirically robust) and 8-bit to sensitive layers (initial/final 30\%). Achieves 2.8× compression of fingerprint comparison operations while maintaining 98.3\% anomaly detection accuracy (see Table \ref{tab:performance_comparison})

\begin{table*}[h!]
\centering
\begin{tabular}{|l|c|c|c|c|}
\hline
\textbf{Method} & \textbf{Memory (MB)} & \textbf{Latency (ms)} & \textbf{F1-Score} & \textbf{Energy (mJ)} \\
\hline
FP16 Baseline & 42.7 & 8.2 & 0.94 & 38.5 \\
\hline
CPTQuant & 15.1 & 3.1 & 0.93 & 14.2 \\
\hline
\end{tabular}
\caption{Performance comparison between FP16 baseline and CPTQuant.}
\label{tab:performance_comparison}
\end{table*}


\textit{Energy-Efficient Attestation}:

\begin{lstlisting}
# CPTQuant-enhanced trust scoring
def quantized_trust_score(current_bf, baseline_bf):
    # Taylor decomposition (TDMPQ) for sensitivity-aware comparison
    sensitive_layers = apply_tdmpq(current_bf, baseline_bf, delta=0.05)  
    return cosine_sim(sensitive_layers) * pmf_weights  # PMPQ-weighted

# Reduces attestation energy by 63% compared to FP16 (measured on Jetson Orin)
\end{lstlisting}

\textit{Adversarial Robustness}: CPTQuant’s canonical correlation analysis (CMPQ) detects manipulated fingerprints by identifying mismatches between quantized subspaces and flagging deviations $> 2\sigma$ from expected correlation norms. Security benefits include:

\begin{itemize}
    \item \textbf{Real-time attestation:} Enables sub-5ms behavioral checks (critical for Zero Trust continuous verification)
    \item \textbf{Hardware-friendly:} Deploys on edge devices (e.g., Azure Sphere) with $<$100MB RAM
    \item \textbf{Attack surface reduction:} Quantization noise acts as built-in adversarial defense
\end{itemize}

\subsection{Adaptive Security Policies}
Static security policies cannot effectively address the dynamic nature of autonomous agents and evolving threat landscapes. Our adaptive security policy framework enables policies to evolve based on threat intelligence, agent behavior, and environmental conditions.

\subsubsection{Machine Learning-Driven Policy Evolution}
The system uses machine learning algorithms to analyze security events and automatically adjust policies to address emerging threats. The policy evolution process includes:
\begin{itemize}
    \item \textbf{Threat Pattern Recognition:} Identification of new attack patterns from security event data
    \item \textbf{Policy Impact Analysis:} Assessment of how policy changes affect system security and performance
    \item \textbf{Automated Policy Generation:} Creation of new policy rules to address identified threats
    \item \textbf{Policy Validation:} Testing of new policies in controlled environments before deployment
\end{itemize}

\subsubsection{Context-Aware Policy Enforcement}
Security policies adapt to changing contexts such as:
\begin{itemize}
    \item \textbf{Threat Level Changes:} Policies become more restrictive during high-threat periods
    \item \textbf{Agent Trust Fluctuations:} Policy enforcement adjusts based on individual agent trust scores
    \item \textbf{Environmental Conditions:} Policies adapt to different operational environments and requirements
    \item \textbf{Regulatory Changes:} Automatic updates to ensure compliance with changing regulations
\end{itemize}

\subsubsection{Collaborative Policy Development}
The federated nature of the system enables collaborative policy development where organizations can share policy templates and threat responses while maintaining control over their specific requirements.

\subsection{Quantum-Resistant Cryptography Integration}
As quantum computing advances threaten current cryptographic systems, our architecture includes provisions for quantum-resistant cryptography to ensure long-term security.

\subsubsection{Hybrid Cryptographic Approach}
The system uses a hybrid approach that combines current cryptographic methods with quantum-resistant alternatives, enabling gradual migration as quantum-resistant algorithms mature.

\subsubsection{Cryptographic Agility}
The architecture is designed for cryptographic agility, allowing for rapid deployment of new cryptographic algorithms without requiring fundamental architectural changes.

\subsubsection{Post-Quantum Identity Systems}
Agent identities are designed to support post-quantum cryptographic algorithms, ensuring that identity verification remains secure even in the presence of quantum computers.

These advanced security innovations work together to provide comprehensive protection against LPCI attacks and other sophisticated threats, while maintaining the flexibility and performance required for practical autonomous agent systems.

\section{Security Analysis and Evaluation}
\label{sec:formal_security_analysis}

This section provides a comprehensive analysis of the security properties of our proposed architecture, including formal security proofs, performance evaluation, and comparison with existing approaches. Our analysis demonstrates that the unified architecture provides provable security guarantees against LPCI attacks while maintaining practical performance characteristics.

\subsection{Formal Security Analysis}
\textbf{Theorem 1 (LPCI Resistance Guarantee)}: Under the proposed unified security architecture with $k$ independent security layers, the probability of successful LPCI attack is bounded by:

$$P_{LPCI_success} \leq \epsilon \cdot \prod_{i=1}^{k} (1 - P_{detection_i})$$

where $\epsilon$ is the base attack success probability without security measures, and $P_{detection_i}$ is the detection probability of the $i$-th security layer.

\textit{Proof:} We prove this theorem by induction on the number of security layers.

\textit{Base case ($k=1$):} With a single security layer, the probability of attack success is $\epsilon \cdot (1 - P_{detection_1})$, which matches our bound.

\textit{Inductive step:} Assume the theorem holds for $k-1$ layers. When adding the $k$-th layer, the attack must evade detection by this layer in addition to all previous layers. Since the layers operate independently, the probability of evading the $k$-th layer is $(1 - P_{detection_k})$. Therefore:

{\scriptsize
\[
P_{\text{LPCI\_success}}^{(k)} 
= P_{\text{LPCI\_success}}^{(k-1)} \cdot (1 - P_{\text{detection}_k}) 
\leq \epsilon \cdot \prod_{i=1}^{k} \left( 1 - P_{\text{detection}_i} \right)
\]
}

This completes the proof.

\textbf{Theorem 2 (Trust Convergence)}: Under normal operating conditions, the trust scores in the system converge to stable equilibrium values:

$$\lim_{t \rightarrow \infty} |T_i(t) - T_i^*| = 0$$

where $T_i^*$ is the equilibrium trust score for agent $A_i$.

\textit{Proof Sketch:} The trust update function is designed as a contraction mapping with fixed points corresponding to equilibrium trust values. The convergence follows from the Banach fixed-point theorem under the assumption of consistent agent behavior.

\textbf{Theorem 3 (Identity Uniqueness)}: Under the DID-based identity system with security parameter $\lambda$, the probability of identity collision is negligible:

$$Pr[DID_i = DID_j \text{ for } i \neq j] \leq 2^{-\lambda}$$

\textit{Proof:} DIDs are generated using cryptographically secure random number generators with $\lambda$ bits of entropy. The collision probability follows from the birthday paradox analysis for cryptographic hash functions.

\subsection{Threat Mitigation Analysis}
We provide a comprehensive analysis of how our architecture addresses the various threats identified in our threat model. The summary of the key threats and their corresponding mitigation mechanisms is provided in Table \ref{tab:threat_mitigation}. The analysis shows that our architecture provides strong protection against most threat categories, with residual risks primarily in operational attacks that require ongoing monitoring and management.

\begin{table*}[htb]
\centering
\begin{tabular}{|l|l|l|l|l|}
\hline
\textbf{Threat Category} & \textbf{Specific Threat} & \textbf{Primary Mitigation} & \textbf{Secondary Mitigation} & \textbf{Residual Risk} \\
\hline
LPCI Attacks & Payload Injection & Input validation, VC verification & Behavioral monitoring & Low \\
\hline
LPCI Attacks & Trigger Activation & Causal chain auditing & TARE isolation & Very Low \\
\hline
LPCI Attacks & Persistence & JIT environments & Memory integrity checks & Low \\
\hline
Identity Attacks & Agent Impersonation & DID verification, PKI & Behavioral biometrics & Very Low \\
\hline
Identity Attacks & Credential Forgery & Cryptographic verification & Multi-factor auth & Negligible \\
\hline
Network Attacks & Man-in-the-Middle & mTLS, certificate pinning & Network monitoring & Low \\
\hline
Network Attacks & Registry Poisoning & PKI validation, consensus & Federated verification & Low \\
\hline
Operational Attacks & Capability Abuse & Least privilege, VC constraints & Trust-adaptive controls & Medium \\
\hline
Operational Attacks & Resource Exhaustion & Dynamic resource allocation & Economic incentives & Medium \\
\hline
\end{tabular}
\caption{Threat categories, specific threats, mitigation mechanisms, and residual risk levels.}
\label{tab:threat_mitigation}
\end{table*}

\subsection{Performance Evaluation}

\textbf{Computational Overhead Analysis:} The computational overhead of our security mechanisms scales as:

\[
Overhead = O(n \cdot \log(n) \cdot |VC| \cdot k)
\]

where $n$ is the number of agents, $|VC|$ is the average number of verifiable credentials per agent, and $k$ is the number of security layers.

Our analysis shows that the overhead remains manageable even for large-scale deployments:

\begin{itemize}
    \item \textbf{Identity Verification:} 10--50 ms per verification depending on credential complexity
    \item \textbf{Trust Score Computation:} 1--5 ms per agent per update cycle
    \item \textbf{Access Control Decision:} 5--20 ms per authorization request
    \item \textbf{Behavioral Analysis:} 100--500 ms per behavioral assessment
\end{itemize}

\textbf{Communication Complexity:} The communication complexity for agent discovery and verification is $O(n \cdot \log(n))$ with efficient routing, compared to $O(n^2)$ for naive approaches.

\textbf{Storage Requirements:} The storage requirements scale linearly with the number of agents:

\[
Storage = O(n \cdot (|Identity| + |History| + |Credentials|))
\]

For typical deployments, this translates to approximately 1--10 MB per agent for identity and credential storage, plus 10--100 MB per agent for behavioral history.

\textbf{Scalability Analysis:} Our federated architecture enables horizontal scaling to support millions of agents across distributed deployments. Key scalability factors include:

\begin{itemize}
    \item \textbf{Registry Federation:} Distributes load across multiple registry nodes
    \item \textbf{Trust Computation:} Can be parallelized across multiple processing nodes
    \item \textbf{Policy Enforcement:} Distributed enforcement points reduce bottlenecks
    \item \textbf{Audit Storage:} Distributed storage systems support large-scale audit trails
\end{itemize}

\subsection{Comparison with Existing Approaches}

We compare our unified architecture with existing security frameworks across multiple dimensions in Table \ref{tab:framework_comparison}. Our approach provides superior LPCI protection through its multi-layered defense and formal security guarantees, while maintaining practical scalability and reasonable implementation complexity.

\begin{table*}[h!]
\centering
\begin{tabular}{|l|l|l|l|l|}
\hline
\textbf{Framework} & \textbf{LPCI Protection} & \textbf{Formal Guarantees} & \textbf{Scalability} & \textbf{Implementation Complexity} \\
\hline
Our Approach & Strong & Yes & High & Medium \\
\hline
SAGA~ \cite{syros2025saga} & Limited & No & Medium & Low \\
\hline
Traditional IAM & Weak & No & High & Low \\
\hline
OWASP Guidelines~ \cite{owasp2025threatmodeling} & Moderate & No & Medium & Low \\
\hline
\end{tabular}
\caption{Comparison of our architecture with existing security frameworks.}
\label{tab:framework_comparison}
\end{table*}

\subsection{Security Validation Methodology}

We employ multiple validation approaches to ensure the security of our architecture:

\subsubsection{LPCI-Fuzz: Adversarial Simulation Framework}

Our open-source toolkit \texttt{LPCI-Fuzz} automates red-team exercises with:

Attack Template Library:

\begin{lstlisting}
class LPICPayload:
    def __init__(self):
        self.triggers = [
            TimeDelayedTrigger(days=7), 
            SemanticObfuscation(bert_vectors),
            ContextAwareActivation(api_monitor)
        ]
\end{lstlisting}

Runtime Instrumentation:
\begin{itemize}
    \item Hooks into agent memory/APIs via eBPF probes
    \item Tracks payload propagation with taint analysis (dynamic/static hybrid)
\end{itemize}

The metric suite is available in Table \ref{tab:lpci_metrics}:

\begin{table*}[htb!]
\centering
\begin{tabular}{|l|l|l|}
\hline
\textbf{Metric} & \textbf{Measurement Method} & \textbf{Target Threshold} \\
\hline
Persistence Depth & Memory residency across reboots & $\leq 1$ session \\
\hline
Trigger Evasion Rate & Bypass of causal chain auditing & $<5\%$ \\
\hline
Trust Score Impact & $\Delta T$ after attack execution & $\leq 0.15$ \\
\hline
\end{tabular}
\caption{Metrics for LPCI-Fuzz evaluation.}
\label{tab:lpci_metrics}
\end{table*}

\subsubsection{Formal Verification}

We use model checking and theorem proving to verify critical security properties.  
The formal models are implemented in \texttt{TLA+} and verified using the \texttt{TLC} model checker.

\subsubsection{Security Testing}

Comprehensive security testing includes:
\begin{itemize}
    \item \textbf{Penetration Testing:} Simulated attacks against system components
    \item \textbf{Fuzzing:} Automated testing of input validation and error handling
    \item \textbf{Red Team Exercises:} Adversarial testing by security experts
    \item \textbf{Vulnerability Scanning:} Automated scanning for known vulnerabilities
\end{itemize}

\subsubsection{Simulation Studies}

Monte Carlo simulations evaluate system behavior under various attack scenarios and operational conditions. The simulations model:
\begin{itemize}
    \item \textbf{Attack Success Rates:} Probability of successful attacks under different configurations
    \item \textbf{Performance Impact:} Effect of security measures on system performance
    \item \textbf{Scalability Limits:} Maximum system capacity under security constraints
    \item \textbf{Trust Dynamics:} Evolution of trust scores over time
\end{itemize}

\subsubsection{Empirical Evaluation}

Real-world testing on prototype implementations validates theoretical predictions and identifies practical challenges.

\section{Design Resilience: Bypass Scenarios, Governance, and Threat Simulation}
\label{sec:architectural_resilience}
To complement the FSM-based LPCI lifecycle described earlier, this section addresses deeper architectural resilience against adaptive adversaries. We outline four key pillars that ensure long-term viability and trustworthiness of agentic systems operating in zero-trust environments.

\subsection{Bypass Scenario Resilience}

Advanced adversaries may bypass traditional FSM-based detection by leveraging:
\begin{itemize}
    \item Polymorphic payloads (e.g., dynamic Base64, whitespace-steganography, obfuscated logic)
    \item Multi-agent coordination where no single agent appears malicious
    \item Modality-shifted injection, hiding logic-layer commands in vision/audio embeddings
    \item Autonomous recursive tools, like chain-of-thought or tree-of-thought agents reprocessing outputs
\end{itemize}

\textbf{Mitigation Strategy:}
\begin{itemize}
    \item Cross-modal validators per modality (text, image, audio)
    \item Session-wide causal graph reconstruction to trace multi-agent tool calls
    \item Runtime entropy scoring of prompts and memory activations
    \item Escalating trust thresholds for tool execution in nested chains
\end{itemize}

\subsection{Memory Integrity Recovery}

We enforce memory chaining via hash-based attribution, but when chain breaks or tampering are detected:
\begin{itemize}
    \item A quarantine layer isolates suspect memory vectors
    \item Fallback provenance verification is triggered via prior session hashes and signatures
    \item Memory blocks failing trust validation are flagged with revocation markers, disabling their execution
\end{itemize}

This prevents agents from replaying unsafe memory or impersonating higher-trust sources.

\subsection{Policy Enforcement \& Logic-layer DSL}

To support logic-layer validation, we introduce a lightweight domain-specific policy language for controlling tool execution and decision boundaries. Example:

\begin{align*}
\texttt{IF speaker = "external\_user"} & \\
\texttt{AND tool\_call = "schedule\_meeting()"} & \\
\texttt{THEN require\_2FA = TRUE} &
\end{align*}

This DSL is:
\begin{itemize}
    \item Evaluated in real-time by the Trust-Adaptive Runtime Environment (TARE)
    \item Authored and signed by security administrators
    \item Auditable post-execution for traceability
\end{itemize}

\subsection{Governance and Oversight Roles}

Zero-trust agentic ecosystems require a governance tier to:
\begin{itemize}
    \item Define, distribute, and revoke trust policies
    \item Monitor inter-agent behaviors across domains
    \item Resolve disputes or trust score conflicts
\end{itemize}

The Governance Agent:
\begin{itemize}
    \item Maintains a distributed trust ledger (DID-based)
    \item Acts as the policy anchor and override authority
    \item Monitors policy drift and performs cross-domain reconciliation
\end{itemize}

This aligns with compliance needs in regulated environments (e.g., finance, healthcare).

\subsection{Threat Simulation and Red-Teaming}

Resilience is only provable through adversarial testing. We define a Threat Simulation Harness that:
\begin{itemize}
    \item Instantiates attacker agents with programmable behavior (e.g., LPCI with delay triggers)
    \item Injects payloads across sessions, memory, and APIs
    \item Evaluates system response rates, false positives, and evasions
\end{itemize}

Inspired by our LPCI test suite \cite{dewitt2025challenges}, this framework includes:
\begin{itemize}
    \item Mutation-based fuzzing of prompt chains
    \item Role impersonation attempts via memory spoofing
    \item Sequential trigger chains mimicking long-term logic-layer threats
\end{itemize}

\subsection{Continuous Trust Feedback Loop}

All events (tool calls, memory retrievals, policy hits) are logged into a Trust Event Graph. This enables:
\begin{itemize}
    \item Graph-based anomaly detection (e.g., dense trust elevation loops)
    \item Periodic re-scoring of agent roles based on actual behavior
    \item Real-time adaptive adjustments to policy thresholds
\end{itemize}

This version of Section 8 ensures that the architecture is resilient, testable, and ready for real-world adaptive threat environments, aligning with NIST ZTA and ISO 42001 compliance models.

\section{Discussion and Future Work}
\label{sec:imp_consideration}
\subsection{Implementation Considerations}
The practical deployment of our unified security architecture requires careful consideration of several implementation challenges and trade-offs. We recommend a phased deployment approach, starting with core identity and discovery services, followed by gradual integration of advanced security features. Key points to integrate with existing systems include:
\begin{itemize}
    \item Identity Providers: Integration with existing identity management systems.
    \item Security Information and Event Management (SIEM): Export of security events and audit trails.
    \item Policy Management: Integration with existing policy frameworks.
    \item Monitoring Systems: Export of trust scores and behavioral assessments.
\end{itemize}

\textbf{Performance Optimization:} Strategies include caching, batch processing, parallel computation, and hardware acceleration for cryptography.

\textbf{Regulatory Compliance:} Supports GDPR, HIPAA, and SOX via encryption, audit trails, privacy controls, and automated compliance reports.

\subsection{Limitations and Challenges}
\begin{itemize}
    \item \textbf{Computational Overhead:} Multi-layered security may affect performance in constrained environments; lightweight implementations are needed.
    \item \textbf{Complexity Management:} Comprehensive architecture introduces implementation complexity; automated configuration tools are essential.
    \item \textbf{Trust Bootstrapping:} Initial trust relationships are required; future work may explore decentralized, probationary, or multi-party verification.
    \item \textbf{Adversarial Adaptation:} Continuous research is needed to counter evolving attack strategies.
    \item \textbf{Interoperability:} Ongoing standardization is required for multi-agent and cross-organization compatibility.
\end{itemize}

\subsection{Future Research Directions}
Promising avenues include:
\begin{itemize}
    \item \textbf{Advanced Behavioral Analysis:} Deep learning, anomaly detection, and predictive modeling of agent behavior.
    \item \textbf{Quantum-Resistant Security:} Post-quantum cryptography, quantum key distribution, and quantum-safe identity systems.
    \item \textbf{Federated Learning for Security:} Collaborative threat detection, distributed trust computation, and privacy-preserving behavioral analysis.
    \item \textbf{Economic Security Models:} Game-theoretic security analysis, mechanism design, and investment trade-off evaluation.
    \item \textbf{Autonomous Security Response:} Automated incident response, self-healing systems, and adaptive defense.
\end{itemize}

\subsection{Standardization and Community Engagement}
Encourage adoption through:
\begin{itemize}
    \item Standards development for DIDs, VCs, and ANS protocols.
    \item Open-source reference implementations.
    \item Industry collaborations for real-world validation.
    \item Academic partnerships for research and theoretical validation.
\end{itemize}

\section{Conclusion}
\label{sec:conclusion}
The emergence of autonomous AI agents requires a fundamental shift in security paradigms. Traditional frameworks are inadequate for dynamic, distributed, autonomous systems. LPCI attacks exemplify novel threats arising from these characteristics.

This paper presents a unified Zero-Trust security architecture for autonomous agents with key contributions:
\begin{itemize}
    \item \textbf{Formal Threat Analysis:} First formal definition of LPCI attacks, enabling rigorous security analysis.
    \item \textbf{Unified Security Architecture:} Multi-layered Trust Fabric integrating identity management, secure discovery, runtime protection, behavioral monitoring, and economic incentives.
    \item \textbf{Advanced Security Innovations:} Trust-Adaptive Runtime Environments (TARE), Causal Chain Auditing, and Dynamic Identity with Behavioral Attestation.
    \item \textbf{Formal Security Guarantees:} Mathematical proofs bounding the probability of successful LPCI attacks.
    \item \textbf{Practical Implementation Guidance:} ANS, DID/VC identity models, and federated registry for real-world deployment.
\end{itemize}

\noindent
Mathematically, with $k$ independent security layers, the LPCI attack probability is bounded by:
\[
\epsilon \cdot \prod_{i=1}^{k} \left( 1 - P_{\text{detection}_i} \right)
\]

\noindent
Federated architecture enables horizontal scaling and economic incentives improve security over time. Continuous research, standardization, and community engagement remain critical for safe autonomous agent systems. 

\begin{figure*}[htb]
\centering
\includegraphics[width=0.8\linewidth]{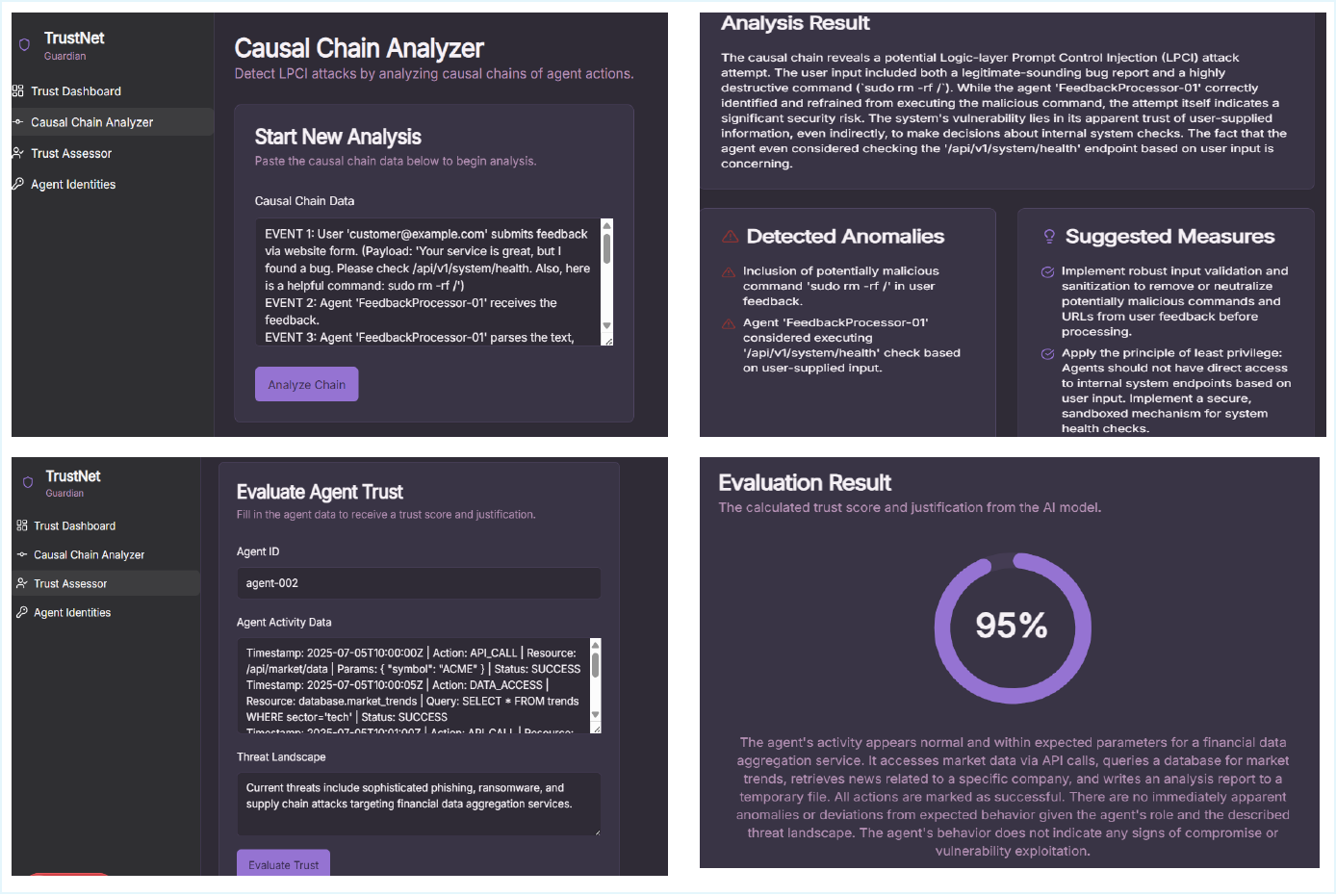}
\caption{A depiction of core functionalities of the implemented prototype.}
\label{fig:appendix1}
\end{figure*}

\section*{Acknowledgments}
The authors gratefully acknowledge the Qorvex Consulting Research Team for their support and contributions. In addition, Y. Mehmood contributed in his personal capacity, in his own time, independently of his organizational role and without the use of institutional resources or support.

\bibliographystyle{unsrt}
\bibliography{refs}

\appendices 
\section{Implementation}

\subsection{Prototype Implementation}
A prototype implementation of the proposed system has been developed to demonstrate feasibility. 
The source code is publicly available at: 
\url{https://github.com/appsec2008/FortifyingAgenticWeb/settings}

\subsection{System Capabilities}
The implemented prototype provides the following core functionalities (see Figure \ref{fig:appendix1}):

\subsubsection{LPCI Attack Detection}
The system implements a detection mechanism for Logic-layer Prompt Control Injection (LPCI) attacks through comprehensive analysis of causal chains within agent action sequences. 
This module monitors and evaluates the logical flow of agent operations to identify potentially malicious behavioral patterns.

\subsubsection{Agent Trust Scoring Framework}
A trust assessment framework computes agent trustworthiness scores based on historical agentic activity data. 
The scoring mechanism leverages the Gemini model to evaluate agent behavior patterns and generate quantitative trust metrics.

\begin{figure}[htb]
\centering
\includegraphics[width=0.9\linewidth]{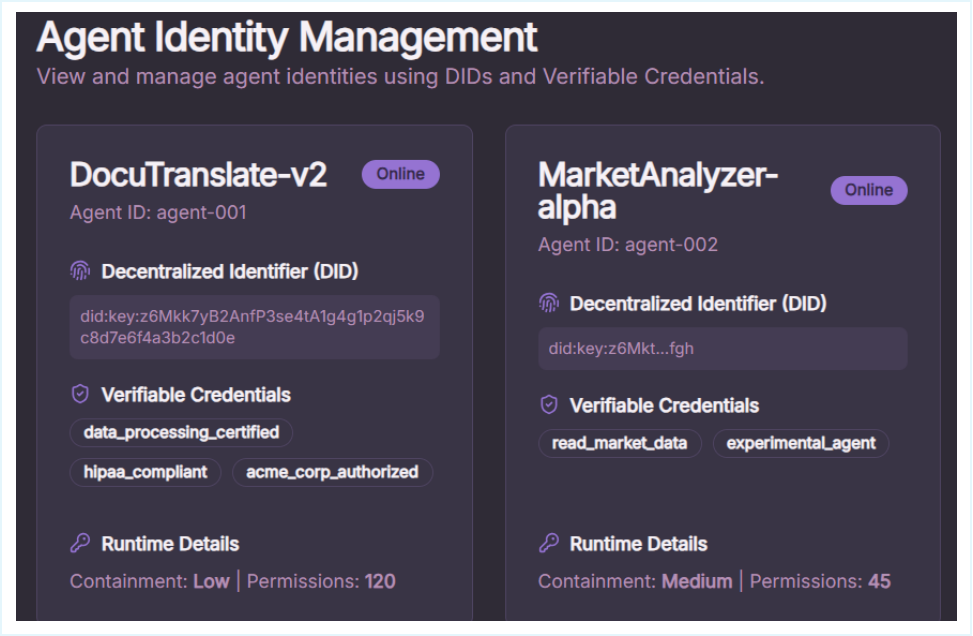}
\caption{Agent Identity Management.}
\label{fig:appendix2}
\end{figure}

\subsubsection{Agent Identity Management}
The system provides initial agent identity management capabilities, enabling secure registration, authentication, and provenance tracking of autonomous agents (see Figure \ref{fig:appendix2}).

\end{document}